# Magnetic, Magnetoelastic and Corrosion Resistant Properties of (Fe–Ni)-Based Metallic Glasses for Structural Health Monitoring Applications

Ariane Sagasti [1,*], Verónica Palomares [2], Jose María Porro [3,4], Iñaki Orúe [5], M. Belén Sánchez-Ilárduya [5], Ana Catarina Lopes [3] and Jon Gutiérrez [1,3]

[1] Deptartment of Electricity and Electronics, Faculty of Science and Technology, Universidad del País Vasco/Euskal Herriko Unibertsitatea, P.O. Box 644, 48080 Bilbao, Spain; jon.gutierrez@ehu.eus
[2] Deptartment of Inorganic Chemistry, Faculty of Science and Technology, Universidad del País Vasco/Euskal Herriko Unibertsitatea, P.O. Box 644, 48080 Bilbao, Spain; veronica.palomares@ehu.eus
[3] BCMaterials (Basque Center for Materials, Applications & Nanostructures), Bldg. Martina Casiano, 3rd. Floor, Barrio Sarriena s/n, 48940 Leioa, Spain; jm.porro@bcmaterials.net (J.M.P.); catarinalopes83@gmail.com (A.C.L.)
[4] Ikerbasque, Basque foundation for Science, 48013 Bilbao, Spain
[5] SGIKER, Universidad del País Vasco/Euskal Herriko Unibertsitatea, 48940 Leioa, Spain; inaki.orue@ehu.eus (I.O.); mbelen.sanchez@ehu.eus (M.B.S.-I.)
* Correspondence: ari.sagasti@hotmail.com

**Abstract:** We have performed a study of the magnetic, magnetoelastic, and corrosion resistance properties of seven different composition magnetoelastic-resonant platforms. For some applications, such as structural health monitoring, these materials must have not only good magnetomechanical properties, but also a high corrosion resistance. In the fabricated metallic glasses of composition $Fe_{73-x}Ni_xCr_5Si_{10}B_{12}$, the Fe/Ni ratio was varied (Fe + Ni = 73% at.) thus changing the magnetic and magnetoelastic properties. A small amount of chromium ($Cr_5$) was added in order to achieve the desired good corrosion resistance. As expected, all the studied properties change with the composition of the samples. Alloys containing a higher amount of Ni than Fe do not show magnetic behavior at room temperature, while iron-rich alloys have demonstrated not only good magnetic properties, but also good magnetoelastic ones, with magnetoelastic coupling coefficient as high as 0.41 for $x = 0$ in the $Fe_{73}Ni_0Cr_5Si_{10}B_{12}$ (the sample containing only Fe but not Ni). Concerning corrosion resistance, we have found a continuous degradation of these properties as the Ni content increases in the composition. Thus, the corrosion potential decreases monotonously from 46.74 mV for the $x = 0$, composition $Fe_{73}Ni_0Cr_5Si_{10}B_{12}$ to −239.47 mV for the $x = 73$, composition $Fe_0Ni_{73}Cr_5Si_{10}B_{12}$.

## 1. Introduction

Wear, fracture, and corrosion are considered to be huge problems in engineering structures, which cause not only enormous economy loss, but also great injuries provoked by the failure of structures and materials. The most striking examples concern the sudden and unexpected falling of concrete-made civil structures like houses or bridges, in which the reinforcing metallic inner structure turn out to be covered with concrete. Thus, this covering of the reinforcing metal makes it impossible to see the true working conditions (see Figure 1).

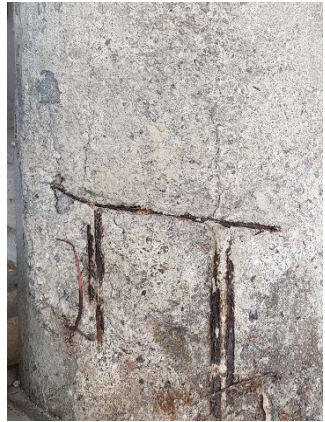

**Figure 1.** Clear evidence of internal corrosion in the reinforcement metal inside a concrete cylindrical column.

In order to prevent that kind of disaster, nowadays structural health monitoring (SHM) studies are performed, which refers to the process of implementing a damage identification strategy for engineering structures [1]. This means to observe a structure or mechanical system over time by taking periodic measurements in order to determine the actual state of the system's health and compare it with the initial undamaged state. Damage caused in the structures could be of different kinds, such as changes related to the materials (for example, in corrosion processes) and/or geometric properties of the system, and defects or flaws. The term damage does not necessarily imply the loss of the structure's functionality; however, the system would not be operating in its optimal way and it could provoke system failures. Nevertheless, if any damage accumulates over long periods of time it can cause fatigue and end with the total damage of the structure.

In order to monitor possible damages inside the structure, amorphous ferromagnetic ribbons (or metallic glasses) are a suitable alternative to develop effective sensors for SHM applications. Metallic glasses fabricated in the form of ribbons can be introduced easily inside the engineering structure during the fabrication process. Due to their magnetic and magnetomechanical properties, these alloys will reflect any change in the structure of the hosting material through a change in the affected magnetic and/or magnetomechanical properties, therefore allowing monitoring and study of the structural system [2,3]. These materials are actually a promising alternative for the current SHM as they present several advantages besides their excellent magnetomechanical properties, for example high corrosion resistance, fabrication low cost, and quick remote detection possibility [4,5]. Such metallic glasses have already been used as sensors and actuators for different applications, showing the wide range of possible sensing parameters [6–8].

The detection operational principle is based on the magnetic and magnetoelastic properties of those materials in which the elastic and magnetic behaviors are coupled, and it can be developed by following two different routes [9]: (a) By using the magnetic susceptibility ($\chi$) of the material as sensing parameter to any applied external stress ($\sigma$). That dependence can be expressed by Equation (1):

$$\frac{d\chi}{d\sigma} = 3\lambda_S \frac{\mu_0 M_S^2}{(2K_u - 3\lambda_S\sigma)^2} \qquad (1)$$

where $M_S$ and $\lambda_S$ are the spontaneous magnetization and saturation magnetostriction of the alloy. Due to their fabrication process, metallic glasses possess null or very low magneto-crystalline anisotropy ($K_u$), and good values of $M_S$ and $\lambda_S$, which directly leads to their high stress sensitivity. The second detection route (b) consists of detecting the changes of the properties of an acoustic wave travelling through the magnetoelastic material. In particular, and for metallic glasses of appropriate form and size, the parameter that is extremely sensitive to different external perturbations is the frequency of the first magnetoelastic resonance mode of the ribbon:

$$f_R = \frac{1}{2L}\sqrt{\frac{E}{\rho}} \tag{2}$$

where now $L$, $E$ and $\rho$ are the ribbon length, Young modulus and density, respectively.

In other words, if the hosting structure (concrete, wood, plastic, etc.) suffers any fracture, wear, or corrosion process, it will cause stress, strain, or a change in the external conditions of the magnetoelastic ribbon embedded in the structure. Due to the magnetoelastic coupling, the metallic glass will react by changing its magnetization state, triggering a magnetic signal that will be detected remotely by a set of coils and through Faraday's electromagnetic induction law, transduced in an electrical signal.

Different compositions of amorphous ferromagnetic magnetoelastic materials have been fabricated and employed to sense and detect different kind of parameters. The composition of the alloys can be modulated in order to look for some specific applications or characteristics of the resonant sensing platform. To fabricate such ribbons in the amorphous phase, which is a metastable state, the composition must be $70-80\%$ of metallic elements ($Fe, Ni, Co, Cr, Au$) and include a $30-20\%$ of metalloids ($B, C, Si, P$, etc., which favors the formation of the amorphous phase) [10,11]. It has been observed that $Fe$-rich amorphous alloys show high strength and hardness, presenting also good magnetic and magnetoelastic properties, low material costs and a superior corrosion resistance, being thus of critical importance for the desired application of the ribbons.

Our previous experience working with metallic glasses exhibiting good corrosion properties, like the Metglas 2826MB3 alloy from VACUUMSCHMELZE GmbH & Co. (Hanau, Germany), told us that even when working within soft saline solutions, clear trends of sample oxidation appeared. Bearing this in mind, in this work we present an extensive magnetic, magnetoelastic, and corrosion resistance study of the family of amorphous ribbons with the nominal composition $Fe_{73-x}Ni_xCr_5Si_{10}B_{12}$ ($0 < x < 73$), where the Fe/Ni ratio was varied from the only Fe-containing sample to the just Ni-containing one. Fe and Ni will be the main elements responsible for the magnetic and magnetoelastic behavior of the ribbons. Nevertheless, not only Ni, but also the presence of other elements in their composition will determine the corrosion resistance of the samples [12–14]. It has been already reported by several authors that the addition of certain elements, such as $Cr$ [15,16] or $Mo$ [17] substantially increases the corrosion resistance of the metallic alloys. In fact, in one of our previous works we observed how the addition of $Cr$ improves the corrosion resistance of these type of magnetoelastic alloys [18]. Bearing this in mind, 5% of $Cr$ was included in our sample's composition. The ultimate goal of the present study is to improve the corrosion resistance of our magnetoelastic ribbons, while maintaining reasonably good magnetic and magnetoelastic properties. This will allow them to be implemented as sensing materials in SHM devices to monitor structural problems of engineering structures. Nevertheless, we have to remark also that other aspects, like the $pH$ of the hosting material (for example concrete has an alkaline $pH$ value) or environment that can alter the corrosion behavior of the metallic glass sensing material, have to be taken into account prior to any implementation in those control devices.

## 2. Materials and Methods

The metallic glasses studied in this work were fabricated at VACUUMSCHMELZE GmbH & Co. KG, Hanau, Germany, in the form of long ribbons using the melt spinning technique, in order to achieve the amorphous state. From the same family, $Fe_{73-x}Ni_xCr_5Si_{10}B_{12}$ ($0 < x < 73$), seven different compositions were prepared, maintaining a small amount of chromium (5% at. of Cr) constant and varying the Fe/Ni ratio (in all cases, Fe + Ni = 73% at.). All the ribbons present a thickness of 25 µm and were cut with a width of 3 mm using a picosecond pulsed laser ablation technology (3D Micromac, microStruct, (Chemnitz, Germany)).

All compositions were fully magnetic and magnetoelastically characterized. Room temperature hysteresis loops were measured by a classical induction method, obtaining therefore the coercive field and saturation magnetization of each strip. The magnetic microstructure of the samples was investigated by using an Evico–Zeiss magneto-optic Kerr effect [19] microscope instrument. The observed changes on the rotation and/or the ellipticity of a linearly polarized beam of light upon reflection from a magnetic surface will depend on the magnetization state of the sample issue of study, therefore allowing for a direct inspection of its magnetic microstructure or magnetic domains [20].

The Curie temperature $(T_C)$ was determined in a Vibrating Sample Magnetometer (VSM) using liquid nitrogen to cool down the samples, and in a Superconducting QUantum Interference Device (SQUID), using liquid helium to reach very low temperatures.

The saturation magnetostriction was determined by using strain gauges KOYWA KFL-02-120-C1-11 (Japan) and an electronic Wheatstone bridge working in half-bridge configuration, including a passive gauge. Magnetoelastic measurements of the $\Delta E$ effect (the change in the Young's modulus with the external applied magnetic field) were carried out using the resonance–antiresonance technique with a home-mounted experimental set-up consisting in three coaxial solenoids in order to apply the bias constant field $(H)$, the alternating field to magnetostrictively excite the sample, and a secondary pick-up coil to monitor the induced magnetization oscillations and detect the frequency of the corresponding magnetoelastic resonance [21].

The corrosion behavior was studied by using the linear potential resistance (LPR) technique with a BioLogic VMP3 Potentiostat/Galvanostat. Measurements were performed in a conventional three-electrode cell, using as working electrode (WE) the fabricated amorphous ribbon, as reference electrode (RE) an Ag/AgCl ingold electrode, and as a counter-electrode (CE) a platinum foil electrode, these last two electrodes purchased from Methrom. We used as electrolyte a saline Phosphate Buffer Saline solution (PBS) 0.01 M purchased from Sigma (Spain) (0.138 M NaCl and 0.0027 M KCl). All the corrosion measurements were made at room temperature (25 °C) and at pH 7.3 [22]. Before measuring the corrosion resistance behavior, the strips were cut in 4 cm-long pieces and carefully cleaned with acetone under sonication for 5 min and dried at room temperature. Samples were left to stabilize for 30 min in the saline solution while measuring the open circuit voltage (OCV), and afterwards we force the working electrode to decrease $-250$ mV from the OCV and scan the potential in the anodic direction at 0.25 mV/s until 250 mV above that OCV. Tafel and $R_P$ fits were performed by analysis of the obtained curves, using the EC-Lab software in order to obtain values for the corrosion potential $(E_{corr})$, the corrosion density $(j_{corr})$, the polarization resistance $(R_P)$ and the Corrosion Rate (CR) [23].

To confirm our observations from electrochemical measurements, X-ray photoelectron spectroscopy (XPS) analyses were performed in a SPECS instrument (Berlin, Germany) that uses monochromatic radiation from $Al\ K\alpha$ (1486.7 eV) and that is provided with a Phoibos 150 1D-DLD analyzer. The system was initially calibrated with Ag ($Ag\ 3d_{5/2}$, 368.26 eV). For each sample, two scans were performed: the first one to determine the elements present in each ribbon (or wide scan, step energy 1 eV, dwell time 0.1 s, pass energy 80 eV) and the second one to carefully analyze each detected element (or detail scan: step energy 0.08 eV, dwell time 0.1 s, pass energy 30 eV). The analyzer is located perpendicular to the sample surface.

## 3. Results

### 3.1. Magnetic and Magnetoelastic Characterization

The first result arising from the magnetic characterization is that, from the seven different compositions fabricated, only four ($x = 0, 12, 24$ and $36.5$, where "$x$" indicates the Ni % at. content) showed ferromagnetic behavior at room temperature. Figure 2a shows the corresponding hysteresis loops, all of them with an almost constant coercivity of $21\ \mu T$. Figure 2b shows the linear decrease of the measured saturation magnetization value with the Ni % at. content. This goes from $\mu_0 M_S = 1.12\ T$, for the $x = 0$ sample (highest 73% at. Fe content, null Ni content), to $\mu_0 M_S = 0.59\ T$ for the $x = 36.5$ one (36.5% Fe content = 36.5% at. Ni content).

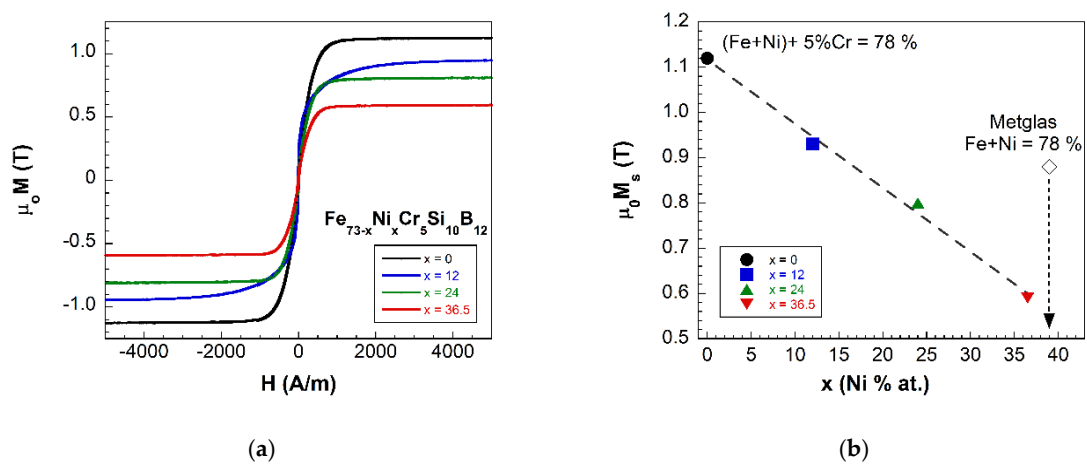

**Figure 2.** (**a**) Hysteresis loops of samples exhibiting ferromagnetic response at room temperature and (**b**) saturation magnetization value decrease as the Ni % at. content increases in the composition (dashed line is a guide to the eye).

To support our previous observations, we proceeded to measure the Curie temperature, $T_C$, of all the fabricated samples by measuring their magnetization versus temperature behavior under an applied magnetic field of $2\,mT$. Obtained curves are shown in Figure 3a, and as it can be observed the samples with high content of Ni ($x = 49, 61$ and $73$) have the Curie temperature below room temperature. Actually, the magnetization of the samples corresponding to a Ni content of 73% cannot be observed, most likely because its Curie temperature is below the measured temperature range. This explains the absence of their hysteresis loops at room temperature. The dependence of the measured Curie temperatures with the ribbon composition is shown in Figure 3b.

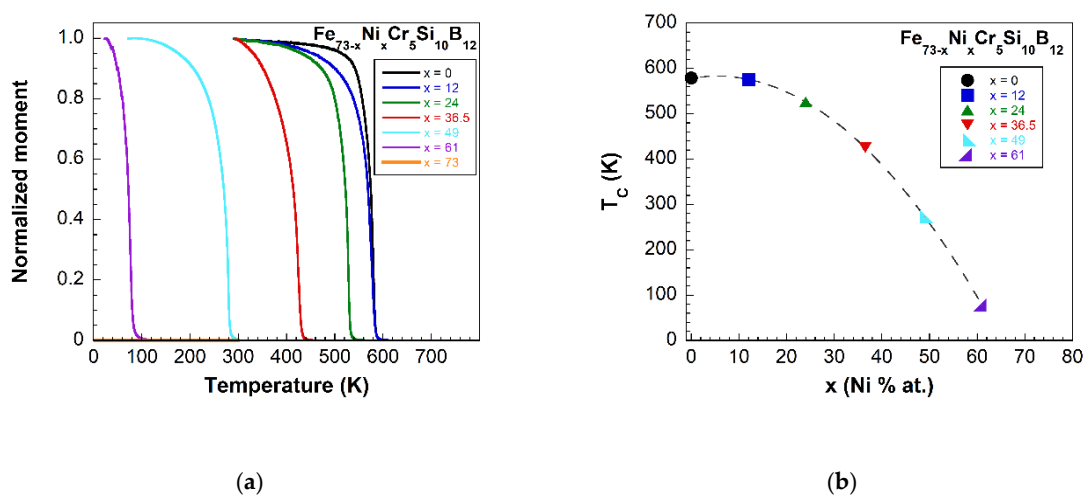

**Figure 3.** (**a**) Variation of the magnetization with the temperature for all the studied samples, and (**b**) Curie temperature dependence with the composition of the samples (dashed line is a guide to the eye).

Afterwards, the samples were characterized by measuring their main magnetoelastic parameters: saturation magnetostriction value ($\lambda_S$) and $\Delta E$ effect. In particular, this last gives an adequate idea of the coupling level between magnetic and elastic properties in these materials, being this responsible for the use as sensors for SHM applications. To quantify it, the so-called magnetoelastic coupling coefficient is used, with the value: $k \approx (\Delta E)_{max}^{1/2}$. Only the samples that show ferromagnetism at room temperature were magnetoelastically characterized. We have observed that as the % at. of Ni increases in the composition, both magnetostriction and $\Delta E$ effect values decrease: from $\lambda_S = 14\ ppm$ and $\Delta E = 17\%$ for the $x = 0$ composition, to $\lambda_S = 2.5\ ppm$ and $\Delta E = 5.6\%$, for the $x = 36.5\%$ at. Ni content sample (see Figures 4 and 5). As a first consequence, the magnetoelastic coupling also decreases from $0.41$ to $0.24$, for the same previous compositions.

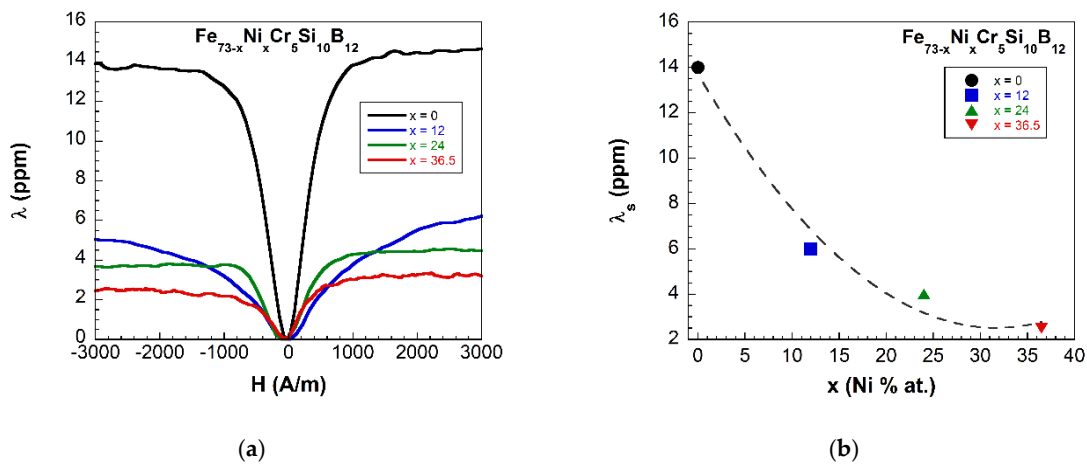

**Figure 4.** (**a**) Measured magnetostriction curves and (**b**) saturation magnetostriction values dependence with the composition of the samples showing ferromagnetism at room temperature (dashed line is a guide to the eye).

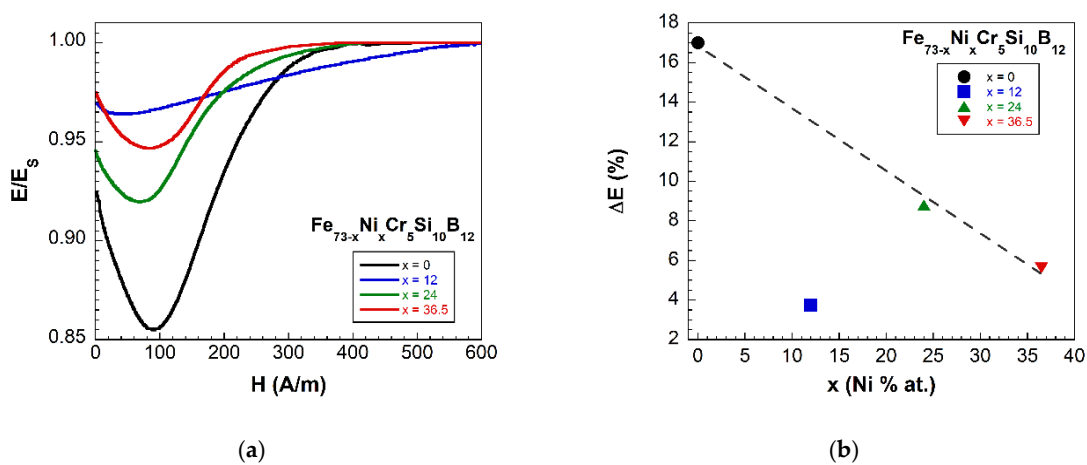

**Figure 5.** (**a**) Change of the Young modulus with the applied magnetic field (or $\Delta E$ effect), and (**b**) measured $\Delta E$ effect values dependence with the composition of the samples showing ferromagnetism at room temperature (dashed line is a guide to the eye).

The drastic drop in the $\Delta E$ effect value for the composition corresponding to $x = 12$ ($Fe_{61}Ni_{12}Cr_5Si_{10}B_{12}$) is noteworthy. This will be extensively discussed in the following sections.

Table 1 summarizes all the magnetic and magnetoelastic parameters measured for all the fabricated samples. For comparison, the properties of the commercially available Metglas 2826MB3 metallic glass, which presents good magnetic and magnetoelastic properties and which is sold as highly corrosion-resistant material, are shown. This alloy, Metglas 2826MB3, has been already reported by several authors to be convenient in order to develop magnetoelastic sensors [6,7,24,25].

**Table 1.** Magnetic and magnetoelastic characterization of all the fabricated $Fe_{73-x}Ni_xCr_5Si_{10}B_{12}$ ($0 < x < 73$) samples. Metglas 2826MB3 alloy-measured magnetic properties are also shown for comparison.

| Composition $Fe_{73-x}Ni_xCr_5Si_{10}B_{12}$ | $\mu_0M_s$ (T) | $T_C$(°C) | $\lambda_s$(ppm) | $\Delta E$ (%) | $k$ |
|---|---|---|---|---|---|
| $x = 0$ | 1.12 | 306 | 14 | 17 | 0.41 |
| $x = 12$ | 0.93 | 302 | 6 | 3.75 | 0.19 |
| $x = 24$ | 0.80 | 253 | 4 | 8.8 | 0.30 |
| $x = 36.5$ | 0.59 | 153 | 2.5 | 5.63 | 0.24 |
| $x = 49$ | – | 0 | – | – | – |
| $x = 61$ | – | −195 | – | – | – |
| $x = 73$ | – | – | – | – | – |
| $Fe_{40}Ni_{38}Mo_4B_{18}$ Metglas® 2826MB3 * | 0.88 | 353 | 11 | 2.5 | 0.16 |

\* Commercially available magnetoelastic ribbon [26,27].

### 3.2. Electrochemical Behavior

Figure 6 shows the measured polarization resistance curves measured against the working electrode potential ($E_{we}$) relative to the Ag/AgCl reference electrode. In all compositions the active area has been of about 1 cm². The corrosion potential value ($E_{corr}$) has been determined by the sharp minima position of each curve. As it can be directly observed, this corrosion potential value shifts towards more positive values as the Fe content in the alloy increases. The lowest value of $E_{corr} = -239.7$ mV has been determined for the non-Fe containing sample, composition $Fe_0Ni_{73}Cr_5Si_{10}B_{12}$.

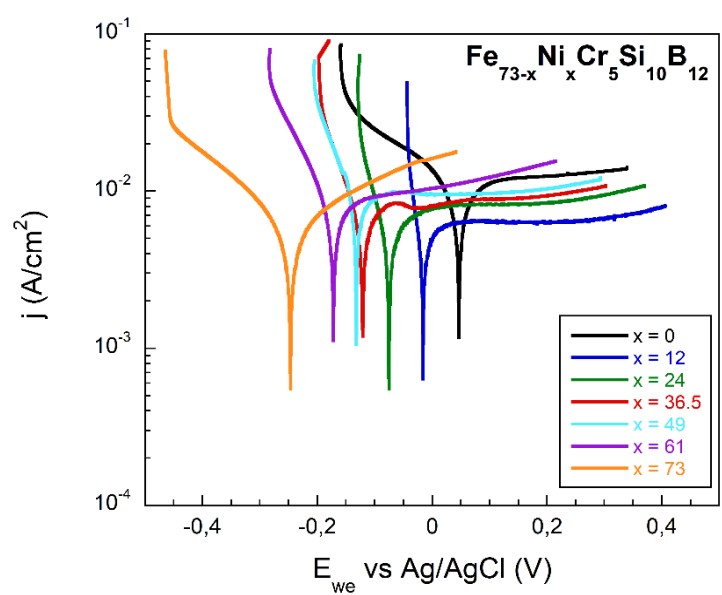

**Figure 6.** Potentiodynamic polarization curves of all the fabricated $Fe_{73-x}Ni_xCr_5Si_{10}B_{12}$ samples, measured in a three-electrode cell in phosphate buffer solution (PBS) at 25 °C and pH 7.3.

From measured LPR curves we performed Tafel and $R_p$ fits in order to obtain current density ($j_{corr}$), polarization resistance ($R_p$) and Corrosion Rate (CR) values for all the fabricated samples. This corrosion rate has to take into account the previously obtained corrosion current, the density of each alloy, and its equivalent weight. Table 2 summarizes all the obtained electrochemical parameters, together with those of the reference sample, Metglas 2826MB3.

**Table 2.** Electrochemical characterization of all the fabricated $Fe_{73-x}Ni_xCr_5Si_{10}B_{12}$ ($0 < x < 73$) samples. Metglas 2826MB3 alloy-measured corrosion properties are also shown for comparison.

| Composition $Fe_{73-x}Ni_xCr_5Si_{10}B_{12}$ | $E_{corr}$ (mV) | $j_{corr}$ (µA/cm²) | $R_p$ ($10^6$ Ohm) | Corrosion Rate (µm/year) |
|---|---|---|---|---|
| $x = 0$ | 47 | 0.013 | 1.05 | 0.035 |
| $x = 12$ | −25 | 0.011 | 1.11 | 0.043 |
| $x = 24$ | −75 | 0.005 | 2.19 | 0.072 |
| $x = 36.5$ | −121 | 0.008 | 1.86 | 0.114 |
| $x = 49$ | −131 | 0.012 | 1.83 | 0.199 |
| $x = 61$ | −172 | 0.011 | 1.86 | 0.197 |
| $x = 73$ | −239 | 0.006 | 2.02 | 0.224 |
| $Fe_{40}Ni_{38}Mo_4B_{18}$ Metglas® 2826MB3 * | −427 | 0.002 | 0.058 | 23.4 |

* Commercially available magnetoelastic ribbon [26,27].

The obtained corrosion rate varies from 0.035 µm/year for the $x = 0$ (only Fe content) to 0.22 µm/year for the $x = 73$ (only Ni content) composition. These results are in good agreement with the change in the previously obtained corrosion potentials, that change from 47 mV to −239 mV for the $x = 0$ (only Fe content) and the $x = 73$ (only Ni content) compositions, respectively.

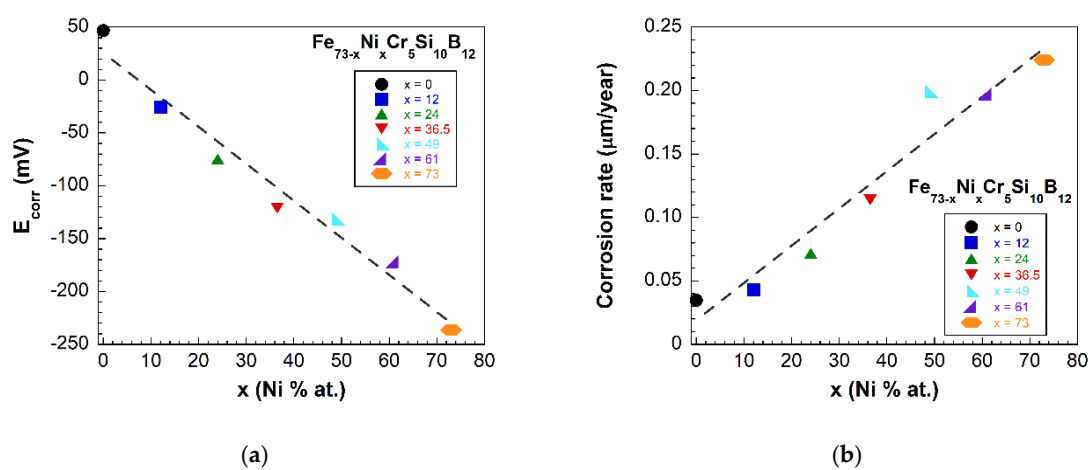

**Figure 7.** (**a**) Corrosion potential ($E_{corr}$) and (**b**) corrosion rate values obtained from Tafel and $R_P$ fittings of the LPR curves obtained for all the fabricated $Fe_{73-x}Ni_xCr_5Si_{10}B_{12}$ samples (dashed lines are a guide to the eye).

Figure 7a shows the obtained $E_{corr}$ values of our fabricated samples. This graph clearly shows the linear decrease of the measured corrosion potential values with the Ni content increase. Finally, in a similar way, Figure 7b shows the calculated corrosion rate values appearing in Table 2, and shows subsequently a linear increase of the obtained corrosion rate values with the Ni content increase.

## 4. Discussion

The magnetic properties of the studied metallic glasses of nominal compositions $Fe_{73-x}Ni_xCr_5Si_{10}B_{12}$ with different Fe/Ni ratios and Fe + Ni = 73% at. show two characteristic features: firstly, from the seven synthesized samples, only the four with a Ni content up to a 36.5% show ferromagnetic behavior at room temperature (see Figure 3). Secondly, a continuous decrease in their measured saturation magnetization values (see Figure 2). The reason for these simultaneous observations arises from the magnetic nature of Fe, Ni and Cr elements.

Fe and Ni are ferromagnetic elements, while Cr is paramagnetic. As expected, the first effect of a 5% at. Cr addition in the composition of our samples is to reduce the magnetization value [28]. This is fully confirmed by comparing the saturation magnetization values of Metglas 2826MB3 and our $Fe_{36.5}Ni_{36.5}Cr_5Si_{10}B_{12}$ sample, the one with the closest composition to the reference sample (see Figure 2b).

The outer electronic structures of Fe and Ni are $3d^64s^2$ and $3d^84s^2$, respectively. The observed decrease in the saturation magnetization values when changing the composition from $Fe_{73}Ni_0$ to $Fe_{36.5}Ni_{36.5}$ can be explained by considering both the energy-band model and the Slater–Pauling curve. To start, the energy-band model [29] tells us that in a ferromagnetic alloy an increase of the number of electrons in the *d* band can enhance or diminish the electron distribution of energy-band, and therefore change the magnetic moment of the alloy. Its experimental observation can be explained by the Slater–Pauling curve [30], which demonstrates that the average magnetic moment of an alloy is a function of the amount of outer electrons. Concerning the elements included in the present study, both Fe and Ni have the 3*d* spin-up band fully occupied, while they have one and three 3*d* spin-down electrons, respectively. Thus, the progressive addition of Ni in the Fe-rich composition alloys enhances the total electron distribution and, as a consequence, gives rise to the observed monotonous decrease in the saturation magnetization values our samples.

Special attention must be payed to the sample $x = 12$, with composition $Fe_{61}Ni_{12}Cr_5Si_{10}B_{12}$. While room temperature hysteresis loops of the $x = 0, 24$ and $36.5$ fabricated samples are almost identical, the hysteresis loop of the sample with $x = 12$ shows that this composition needs more applied magnetic field in order to reach magnetic saturation (in other words, the approach to saturation in the hysteresis loop is more "rounded"; see Figure 8a). Also, the measured $\Delta E$ effect magnitude for this composition shows an abrupt decrease if compared with the other compositions (see Figure 5b). In order to elucidate the origin of such anomalous behavior we have inspected the domain sizes of samples $x = 0$ and $12$ by means of Kerr microscopy.

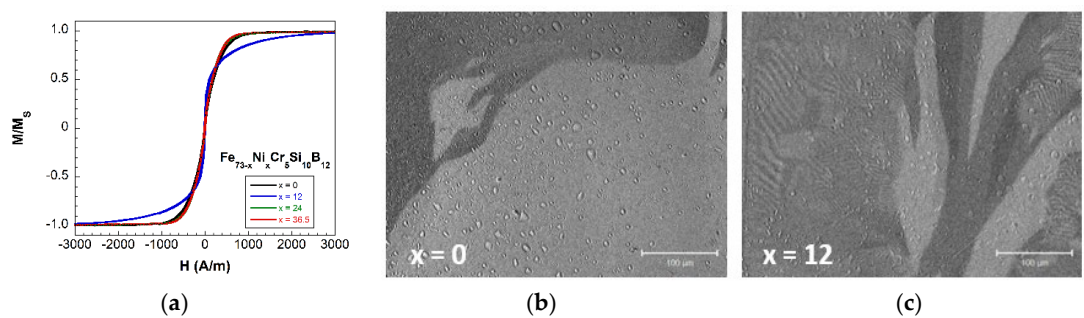

**Figure 8.** (**a**) Normalized hysteresis loops and Kerr microscopy images of samples: (**b**) $x = 0$ and (**c**) $x = 12$ at remanence. Black contrast indicates negative saturation along the field direction, while white contrast corresponds to positive saturation along the field direction.

A direct inspection of Figure 8b,c shows that at remanence, the sample with $x = 0$ shows only big magnetic domains while sample $x = 12$ presents a coexistence of small and large sized magnetic domains. The relation between the domain sizes, and the coercive and saturating fields in patterned [31] and thin film [32] soft magnetic materials, is already well established. The $x = 12$ sample undergoes a reversal process involving sequential flipping of domains, starting from the small domains being flipped upon application of an external field, and then dragging the bigger domains upon reversal of the smaller ones. The fingerprint of this process is precisely the distinct hysteresis loop shape shown in panel (a) of Figure 8, in opposition to a sharp reversal process of the big domains present in the rest of the samples, which results in sharper hysteresis loop shapes. This difference in the bulk magnetization process together with the different inner microstructure of those magnetic domains within the sample, give rise to the observed anomalous behavior in both hysteresis loop and Young's modulus variation one.

Concerning the electrochemical behavior of the samples, our results, shown in Figures 6 and 7 and summarized in Table 2, clearly indicate that $(Fe - Ni)$-based metallic glasses with a small percentage of chromium in their composition exhibit excellent corrosion resistance properties. As expected, the most corrosion resistant alloy is also the one with the lowest corrosion rate $(Fe_{73}Ni_0Cr_5Si_{10}B_{12}, x = 0)$. This is fully justified by the measured high corrosion potential and low current density determined from our measurements [18,33].

Also, we have observed that the gradual addition of Ni to the initial $(Fe_{73}Ni_0Cr_5Si_{10}B_{12}, x = 0)$ sample composition, gives rise to a smooth, but continuous, degrading of those properties. The first reason for such behavior has to be found in the structural characteristics of the amorphous state. This is a homogeneous single phase arisen from the extremely rapid cooling rates necessary to form the glassy state. So, there are no grain boundaries nor dislocations that provide an initial site for corrosion [12,34]. This amorphous structure is usually accompanied by a chemical homogeneity that, if including the proper elements, gives excellent corrosion properties to the amorphous material. This is the case of our Fe − Ni − Cr containing metallic glasses [12,15,34].

The homogeneity in chemical composition favors the appearance of an amorphous oxide film on their surface: the so-called "passivation layer", that retards the charge transport responsible for corrosion phenomenon. The formation of this passive layer occurs very quickly. To gain a deeper insight into the existence of such protective layers onto our amorphous strips, we have performed X-ray photoelectron spectroscopy (XPS) analysis with the $x = 0, 24$ and 49% at. Ni content compositions. Figure 9a–e show the spectra measured for Fe, Ni, Cr, Si and B, respectively, for the three studied compositions.

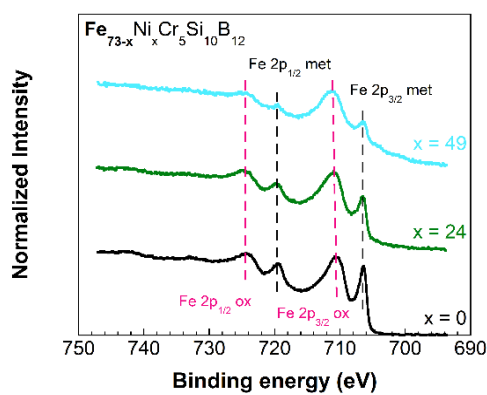

(**a**)

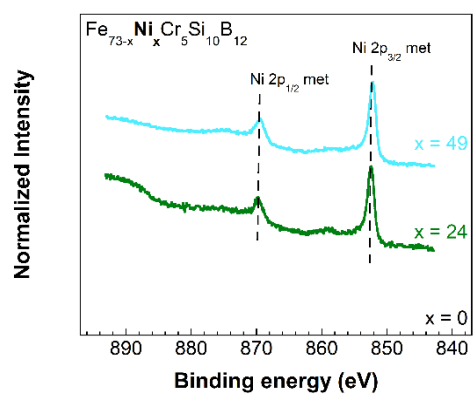

(**b**)

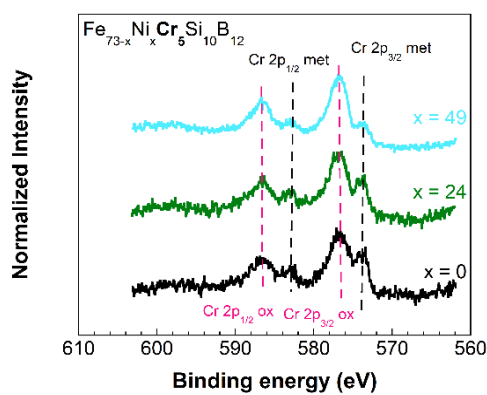

(**c**)

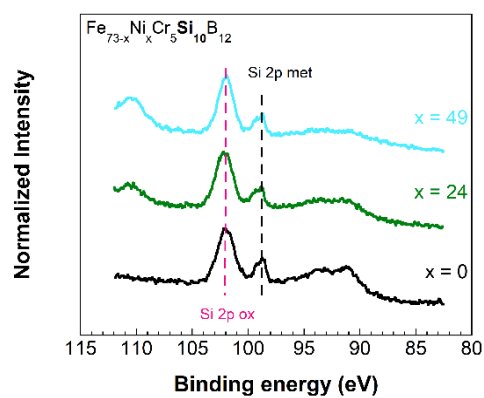

(**d**)

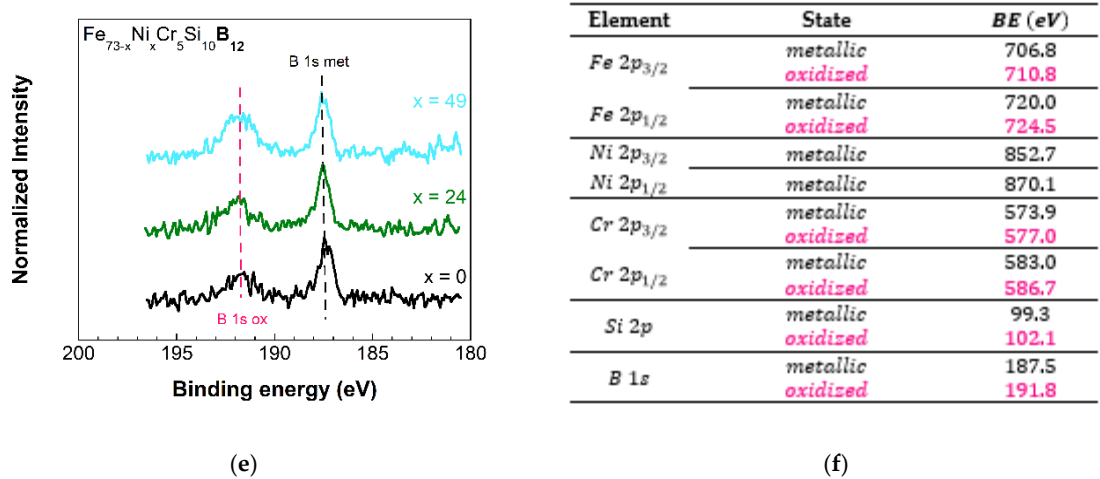

(e)    (f)

**Figure 9.** X-ray photoelectron spectroscopy (XPS) spectra obtained for Fe (**a**), Ni (**b**), Cr (**c**), Si (**d**) and B (**e**) for the three studied samples with compositions $Fe_{73-x}Ni_xCr_5Si_{10}B_{12}$, for $x = 0, 24$ and $49\%$ at. Ni content compositions, and (**f**) binding energy (BE) values for the most representative peaks of each element and state (metallic or oxidized).

Regarding the iron spectra, elemental (metallic) $Fe\ 2p_{3/2}$ and $2p_{1/2}$ emissions located at $706.8$ eV and $720.0$ eV binding energies are observed. Also, $Fe\ 2p_{3/2}$ and $2p_{1/2}$ emissions located at $710.8$ eV and $724.5$ eV are observed, which correspond to oxidized iron (being $FeO$, $Fe_2O_3$ or $Fe_3O_4$ possible oxides) [35].

In the case of nickel, $Ni\ 2p_{3/2}$ and $2p_{1/2}$ elemental (metallic) emissions located around $852.7$ eV and $870.1$ eV binding energies are observed. The presence of oxidized nickel is not clear, since $Ni\ 2p_{3/2}$ and $2p_{1/2}$ emissions located around $858$ eV and $875$ eV are observed, but these emission features could correspond to plasmon loss leaks for metallic Ni.

Related to chromium, $Cr\ 2p_{3/2}$ and $2p_{1/2}$ elemental (metallic) emissions located at $573.9$ eV and $583.0$ eV binding energies are observed. Also, $Cr\ 2p_{3/2}$ and $2p_{1/2}$ emissions located at $577.0$ eV and $586.7$ eV are observed, which correspond to oxidized chromium (probably $Cr_2O_3$) [36].

In the silicon spectra, $Si\ 2p$ elemental (metallic) emission located at $99.3$ eV and $Si\ 2p$ (oxidized) emission located at $102.1$ eV are observed, most probably due to the presence of $SiO_2$.

Finally, the boron spectra show $B\ 1s$ metallic emission located around $187.5$ eV and oxides emission around $191.8$ eV, this last corresponding to oxidized species, as $B_2O_3$ or $FeBO_3$ [37–39].

In all these detected peaks, we assume those elemental or metallic emissions to arise from the bulk alloy. However, all the emissions arising from oxidized species belong to the presence of formed passivation layers that give the corrosion properties to the different alloy compositions. From the quantification of the measured XPS spectra, it is clearly shown that the Fe oxide layer atomic percentage decreases ($6.8\%$ for the $x = 0$ sample, $5.5\%$ for the $x = 24$ sample and $4.5\%$ for the $x = 49$ sample) as the Ni content in the sample increases ($2.2\%$ for the $x = 24$ sample and $5.1\%$ for the $x = 49$ sample). The Cr oxide passivation layer slightly changes, $1.2\%$ for the $x = 0$ sample, $1.3\%$ for the $x = 24$ sample and $2.1\%$ for the $x = 49$ sample. More surprisingly, the presence of a relevant Si oxide layer has been confirmed by our measurements, with values of $11.8\%$ for the $x = 0$ sample, $12.2\%$ for the $x = 24$ sample and $14\%$ for the $x = 49$ sample. In summary, in the existing passivation layers the Fe oxide amount decreases as the Ni content in the composition of the sample increases, Cr oxide content practically does not change and there is a significant amount of Si oxide (about $12 - 14\%$) that strongly contributes to the corrosion resistance behavior of our studied samples. These atomic percentage values obtained from the XPS analysis reflect the expected concentration trend in the different samples, but they do not correspond with the theoretical values due to the presence of carbon (usually detected in samples analyzed by XPS) and an excess of oxygen because of the air exposure.

## 5. Conclusions

In the search of innovative materials for Structural Health Monitoring applications, we have shown that metallic glasses of composition $Fe_{73-x}Ni_xCr_5Si_{10}B_{12}$, with $x < 36.5$ show excellent magnetic, magnetoelastic and corrosion resistance properties that present them as ideal candidates for the aforementioned applications.

They are ferromagnetic at room temperature and present good saturation magnetization values (from $0.59\,T$ for $x = 36.5$, to $1.12\,T$ for $x = 0$) as well as magnetoelastic coupling coefficient (from $0.24$ for $x = 36.5$, to $0.41$ for $x = 0$), this last directly related to the feasibility of the materials for remote interrogation. In an analogous way, our electrochemical extensive study has also demonstrated that the same compositions with $x < 36.5$ show the best corrosion resistance behavior with the highest measured values of corrosion potential and polarization resistance combined with the lowest corrosion current densities and corrosion rate (less than $0.2\,\mu m/year$).

To account for such outstanding magnetic and corrosion resistance behavior, each element within the composition of these metallic glasses plays its role. Not only is Fe the element with the strongest magnetic contribution in this family of compositions; the stability of its amorphous state and also the formation of a clear oxide passive layer onto the studied strip surfaces give them excellent corrosion resistance properties. In the case of Ni, its mixture with Fe atoms decrease the measured net magnetization value in all the Fe − Ni-containing samples. Cr (present in a low 5% at. content in all the studied compositions) decreases systematically the saturation magnetization of all the compositions when compared with the one of Metglas 2826 MB3 (the commercial sample also studied and presented for comparison) and also forms a passive protective layer appeared in the same percentage (in all the studied samples). Finally, we have been able to determine the presence of a relevant Si oxide passive layer that also contributes strongly to the excellent corrosion behavior exhibited by our studied compositions.

**Author Contributions:** A.S. and J.G. conceived and designed the work and needed measurements; A.C.L. supervised the fabrication of the samples; A.S., V.P., J.M.P., I.O. and M.B.S.-I. performed the experiments; A.S., A.C.L. and J.G. analyzed the data; A.S. and J.G. wrote the manuscript. All authors discussed the results and implications, and commented on the manuscript at all stages. All authors have read and agreed to the published version of the manuscript.

**Funding:** The authors acknowledge financial support from the Basque Government under ACTIMAT project (KK-2018/00099, Elkartek program) and University Basque Research Groups Funding (IT1245-19). J.M.P. and A.C.L. acknowledge funding from the H2020 Excellent Science-Marie Sklodowska-Curie Actions with the individual fellows no.753025 and no. 701852, respectively.

**Acknowledgments:** Technical and human support provided by SGIker (UPV/EHU, MICINN, GV/EJ, ESF) is gratefully acknowledged. Ana Catarina Lopes also acknowledges Christian Polak from VACUUMSCHMELZE GmbH & Co. KG, Germany for supplying the studied alloys.

**Conflicts of Interest:** The authors declare no conflict of interest.